\documentclass[epsf,graphics,usedcolumn]{mn2e}
\usepackage{general_cite,epsf,graphics,graphicx}
\usepackage[T1]{fontenc}
\usepackage{ae,aecompl}

\def\h{$^{\rm h}$}
\def\m{$^{\rm m}$}

\def\deg{\hbox{$^\circ\;$}}

\title{OH megamasers, starburst and AGN activity in Markarian 231} \author[A. M. S. Richards et al.]
      {A. M. S. Richards$^1$\thanks{email: {\tt amsr@jb.man.ac.uk}}, J. H. Knapen$^2$, J. A. Yates$^{3}$,
	R. J. Cohen$^{1}$, J. L. Collett$^{2}$, \newauthor
	M. M. Wright$^{4}$, M. D. Gray$^{5}$ and D. Field$^{6}$ \\ 
$^1$ Jodrell Bank Observatory, The University of
Manchester, Jodrell Bank, Cheshire, SK11 9DL, UK.\\  
$^2$Centre for Astrophysical Research,  University of Hertfordshire,
Hatfield, AL10 9AB, UK.\\ 
$^3$ Department of Physics and Astronomy, University College London,
Gower Street, London WC1E 6BT, UK.\\ 
$^4$ Department of Physics, University
of Bristol, Tyndall Avenue, Bristol, BS8 1TL, UK.\\ $^5$ Astrophysics Group, The University of Manchester, Sackville Street, P.O. Box 88, Manchester M60 1QD, UK \\ $^6$ Intsitute of
Physics and Astronomy, University of Aarhus, DK-8000, Aarhus C,
Denmark.\\
}

\date{Accepted ...
      Received ...
      in original form ...}
\pagerange{000--000}

\begin{document}

\maketitle

\begin{abstract}

We present MERLIN observations of OH maser and radio continuum
emission within a few hundred pc of the core of the Ultra Luminous IR
Galaxy Markarian 231.  This is the only known OH megamaser galaxy
classed as a Seyfert 1.  Maser emission is identified with the 1665-
and 1667-MHz transitions over a velocity extent of 720 km~s$^{-1}$.
Both lines show a similar position-velocity structure including a
gradient of 1.7 km s$^{-1}$ pc$^{-1}$ from NW to SE along the 420-pc
major axis. The (unresolved) inner few tens of pc possess a much
steeper velocity gradient. The maser distribution is modelled as a
torus rotating about an axis inclined at $\sim45$\deg to the plane of
the sky.  We estimate the enclosed mass density to be $320\pm90$
M$_{\odot}$ pc$^{-3}$ in a flattened distribution. This includes a
central unresolved mass of $\la8\times10^{6}$ M$_{\odot}$. All the
maser emission is projected against a region with a radio continuum
brightness temperature $\ge10^5$ K, giving a maser gain of
$\le2.2$. The 1667:1665-MHz line ratio is close to 1.8 (the value
predicted for thermal emission) consistent with radiatively pumped,
unsaturated masers. This behaviour and the kinematics of the torus
suggest that the size of individual masing regions is in the range
$0.25-4$ pc with a covering factor close to unity.  There are no very
bright compact masers, in contrast to galaxies such as the Seyfert 2
Markarian 273 where the masing torus is viewed nearer edge-on.  The
comparatively modest maser amplification seen from Markarian 231 is
consistent with its classification in the unification scheme for
Seyfert galaxies.  Most of the radio continuum emission on 50-500 pc
scales is probably of starburst origin but the compact peak is $0.4$
per cent polarized by a magnetic field running north-south, similar to
the jet direction on these scales. There is no close correlation
between maser and continuum intensity, suggesting that much of the
radio continuum must originate in the foreground and indeed the
relative continuum brightness is slightly greater in the direction of
the approaching jet.  Comparisons with other data show that the jet
changes direction close the nucleus and suggest that the sub-kpc disc
hosting the masers and starburst activity is severely warped.

\end{abstract}

\section{INTRODUCTION}
\label{intro}
Ultraluminous infrared galaxies (ULIRGs) were first discovered using
{\em IRAS} (see review by \pcite{Sanders96}). They are defined as
having luminosities $L_{\rm IR}\ge 10^{12}\,{\rm L}_{\odot}$,.  They
are thought to be powered mainly by bursts of extreme star formation,
with contributions from active nuclei in at least some cases
(e.g. \pcite{Genzel98}). Seemingly without exception, and in contrast
to lower-luminosity starbursts, ULIRGs show evidence for dramatic
intergalactic collisions or mergers when observed in the optical or
NIR (\pcite{Sanders96}; \pcite{Knapen04}).  At such extreme
luminosities, ULIRGS would be detectable out to redshifts $\gg1$ by
new IR telescopes such as {\em Spitzer}.  The galaxy interaction rate
is expected to be greater at larger look-back times
\cite{Schweizer98}.
The
study of these rare objects in the local universe is crucial to our
understanding of galaxy interactions and the triggering of
star-formation at all epochs and, by implication, the development of
structure in the early universe \cite{Kneib04}.

\scite{Sanders88} proposed that ULIRGs are part of a sequence in which
spiral galaxies merge, evolve through a starburst phase and become
optically visible QSOs.  They postulate that gas is
funnelled into the merging nuclei causing nuclear starbursts and the
formation of a self-gravitating nuclear disc on a scale of 1~kpc.  The
disc is thought to fuel the activity of the nucleus. Furthermore,
starburst activity produces enough fast-evolving stars to give rise to
a high rate of mass loss back into the interstellar medium (ISM)
leading to sustained fuelling of the central black hole
\cite{Norman88}.  The merger activity produces a warm ($\sim50$ K)
dusty environment in the nuclear region, providing good conditions for
OH megamaser activity \cite{Randell95}.

Markarian (Mrk) 231 is one of the most luminous and best studied IRAS
galaxies, with $\log (L_{\rm FIR}/{\rm L}_\odot)=12.04$. It has a
redshift of 0.042, yielding an estimated distance of
170~Mpc\footnote{We adopt $H_0=75\,$km\,s$^{-1}$\,Mpc$^{-1}$.  At the
distance of Markarian~231 one arcsec corresponds to 820\,pc.}.
Optical spectroscopy shows broad lines characteristic of a Seyfert 1
nucleus \cite{Boksenberg77}.  A radio jet is apparent in VLBI and VLA
images on scales of sub-pc to hundreds of pc, with a weak counter-jet
(\pcite{Ulvestad99a} and references therein), providing unusually
strong evidence for the presence of an AGN in an OH megamaser region.
The IR luminosity has been partly attributed to starburst activity
\cite{Downes98} and partly to the AGN \cite{Soifer00}, discussed
further in Section~\ref{polarization}. On scales $\la100$ pc the
radio-continuum emission is AGN dominated \cite{Lonsdale03}.  On
greater than kpc scales it is starburst-dominated although the
contribution from the dissipated radio jet may be significant
\cite{Ulvestad99sa}. The spectral indices of the intermediate-scale
radio continuum suggest a diffuse component powered by a disc or halo
of star formation as well as fragmented jet emission to the south
\cite{Taylor99}.


 Mrk 231 shows powerful OH megamaser emission first detected by
\scite{Baan85}. \scite{Baan92} reported emission in all 4 ground-state
transitions; the 1667-MHz line was stronger than the 1665-, 1720- and
1612-MHz lines by factors of approximately 2.3, 13 and 30,
respectively.  Using the Lovell telescope with a beam size of 10
arcmin, \scite{Stavely-Smith87} measured emission over 6.9 MHz and
estimated the total velocity width of the 1667-MHz emission to be
$760\pm100$ km~s$^{-1}$ with a peak maser flux density of 48
mJy. \scite{Klockner03} obtained similar results using the Westerbork
array at a resolution of 14 arcsec, but the European VLBI Network
(EVN) at 39 milli-arcsec (mas) resolution only detected just over half
the OH emission within a region of maximum angular size 150 mas.

The Multi Element Radio Linked Interferometer Network (MERLIN) is
sensitive to 1.4--1.6 GHz emission on scales up to $2$ arcsec (1.6
kpc) with a resolution of $100-200$ mas.  We describe our observations
using MERLIN in Section~\ref{obs}.  We present the continuum and line
results in Section~\ref{results}.  In Section~\ref{dynamic} we explain
the radio morphology and polarization and analyse the maser kinematics
in comparison with other observations of the galaxy on various scales,
which provides limits on the size of masing clouds and the enclosed
mass.  In Section~\ref{maser} we analyse our findings and information
from the literature in order to estimate the maser optical depths and
other characteristics, further constraining the size of masing regions
and their number density.  We make some comparisons with other
megamaser galaxies in Section~\ref{comparisons} and summarise our
conclusions in Section~\ref{conclusions}.

\section{OBSERVATIONS}
\label{obs}

\subsection{Continuum observations}
\label{contobs}

We observed Mrk 231 at 1658 MHz on 1997 June 20 using 6 antennas of
MERLIN for 10 hr with a 16-MHz bandwidth. We also retrieved MERLIN
archive observations at 1420 and 1658 MHz made on 1993 May 23-25 for
$\approx7$ hr at each frequency, using 8 antennas including the Lovell
and Wardle telescopes, giving improved sensitivity on angular scales
up to 6 arcsec. J1302+578 was used as the phase-reference source,
observed alternately with Mrk 231 in a 10-12-min cycle. Its position
is known to better than 1~mas \cite{Ma98} and it is only 1\fdg3 from
Mrk 231. Details of the sources used for calibration of the bandpass
and the flux scale (by comparision with with 3C286, \pcite{Baars77})
are given in Table~\ref{flux}.  The data were reduced using standard
MERLIN procedures \cite{Diamond02} and the {\sc aips} package. For
each epoch, we self-calibrated J1302+578 and interpolated the phase
and amplitude solutions over the Mrk 231 data. We also corrected for
polarization leakage and polarization angle offset (with respect to
3C286). The pointing position for Mrk 231 (referred to hereafter as
the reference position) was 12\h~56\m14\fs2383
+56\degr~52\arcmin~25\farcs210 (J2000).  The absolute position
uncertainty of our images is about 10 mas, taking into account the
errors in the positions of the antennas and J1302+578 and in
transferring the phase corrections to Mrk 231.

Table~\ref{flux} gives the properties of images made using the same
restoring beam at all epochs/frequencies.  This shows that the peak
flux densities are the same within the flux scale uncertainy, but the
total flux detected in 1997 is only 86 per cent of that detected in
1993. The 1993 1658-MHz data were imaged using natural weighting to give
optimum fidelity and sensitivity in studying extended continuum
emission.  The 1997 data were imaged at the same resolution as the
masers for the comparison in Section~\ref{gain}.
\begin{table}
\caption{Properties of the flux calibration source used for each
  epoch/frequency (columns 3 and 4, accurate
to 2--3 per cent) and the peak, noise and
  total flux density of Mrk 231 using a 200-mas circular restoring
  beam (columns 5-7).}
\begin{tabular}{lcclccc}
\hline
Epoch & Freq. &\multicolumn{2}{c}{Flux cal.} & Peak & $\sigma_{\rm rms}$ & Total \\
(yr)  & (MHz) & (source) & (Jy) & \multicolumn{2}{c}{(mJy beam$^{-1}$)} &
  (mJy) \\
(1) & (2) & (3) &(4)&(5)&(6)&(7)\\
\hline
1993 & 1420 & OQ208 & 0.935 & 177.93 & 0.07 & 266.95\\
1993 & 1658 & OQ208 & 1.187 & 175.14 & 0.06 & 265.04 \\ 
1997 & 1658 & 2134+004& 4.9 & 173.00 & 0.07 & 229.24 \\
\hline
\end{tabular}
\label{flux}
\end{table}


\subsection{Maser line observations}
\label{lineobs}
We observed Mrk 231 in spectral line mode on 1997 June 20, using the
 same calibration sources and initial procedures as for the
 contemporaneous continuum observations (Section~\ref{contobs}).  We
 used a bandwidth of 8~MHz divided into 64 frequency channels, giving
 a channel width of 22.5 km~s$^{-1}$.  Mrk 231 has a velocity relative
 to the Local Standard of Rest, $V_{\rm LSR}$, of 12137 km~s$^{-1}$
 using the radio convention in which the fractional shift in line
 frequency is proportional to the recession velocity as a fraction of
 $c$. This corresponds to 12639 km~s$^{-1}$ in the optical
 heliocentric convention ($V_{\rm hel}$ at the epoch of observation.
 Mrk 231 was observed alternately at frequencies corresponding to the
 redshifted 1665- and 1667-MHz OH mainline maser transitions.
 \scite{Yates00} describes the additional data reduction stages for
 MERLIN megamaser observations.

\begin{figure}
\resizebox{8.5cm}{!}{\rotatebox{-90}
{\epsfbox{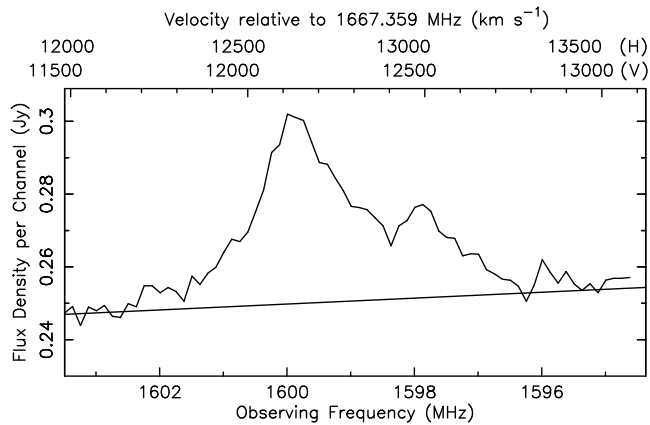}}}
\caption{Scalar average flux density of all emission from Mrk 231 as a
  function of frequency,
  measured by MERLIN in 1997.  The sloping line shows a linear fit to the
  first 18 and last 9 channels which appear to be continuum-only. The
  velocities are given in the  radio LSR convention (V)
  and in the optical heliocentric convention (H). The velocity is
  calculated for the rest frequency of the 1667-MHz line; subtract 338
  km s$^{-1}$ for velocities  with respect to
  the 1665-MHz line.}
\label{VELPROF.PS}
\end{figure}

\begin{figure}
\hspace*{-0.5cm}
\resizebox{9cm}{!}
{\epsfbox{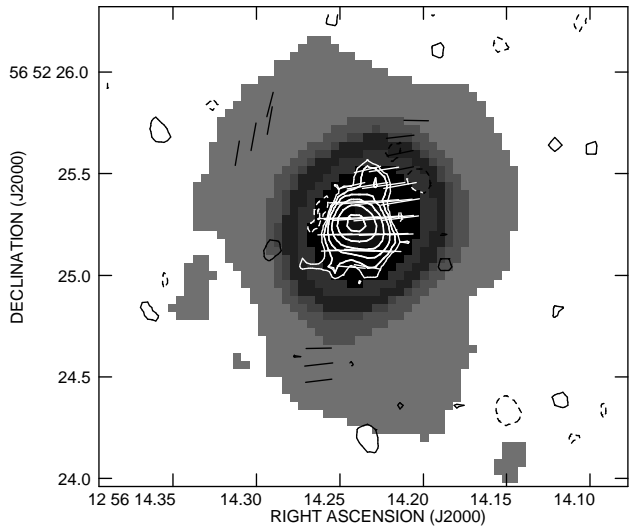}} 
\caption{The greyscale shows the continuum flux at 1658 MHz from Mrk
  231, from the 3$\sigma_{\rm rms}$ level 0.25 mJy beam$^{-1}$ to 200
  mJy beam$^{-1}$.  A ($362 \times 284$) mas$^2$ restoring beam was
  used. The short lines show the linear polarization vectors. 100 mas
  represents 0.25 mJy beam$^{-1}$ polarized intensity.  The contours
  show the sum of all OH maser emission with levels (--1, 1, 2, 4 ...)
  $\times$ 1.2 Jy beam$^{-1}$ km s$^{-1}$ (equivalent to 1.125 mJy
  beam$^{-1}$ in a single channel), using a 120-mas circular restoring beam. }
\label{M+C.GPS}
\end{figure}

\begin{figure}
\hspace*{-0.5cm}
\resizebox{9cm}{!}{\rotatebox{0}
{\epsfbox{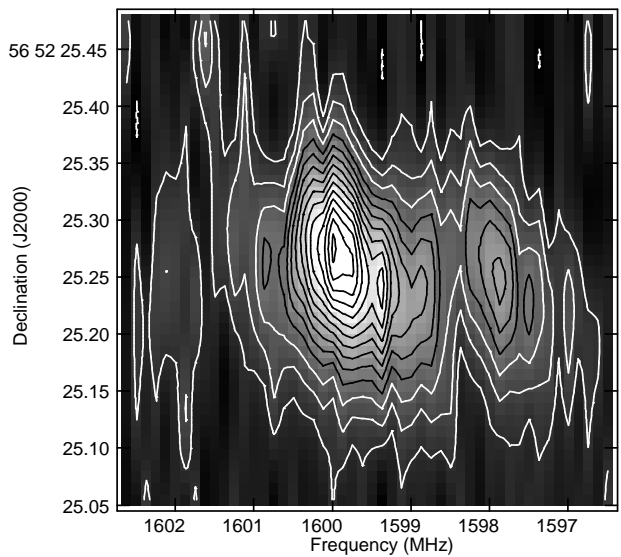}}}
\caption{All maser emission from Mrk 231 summed across the emission
  region along the right ascension axis.  The contours are at
  intervals of 6 per cent of the peak. See
  Fig.~\protect{\ref{VELPROF.PS}} for the velocities of each line as a
  function of frequency.  }
\label{FreqDec.GPS}
\end{figure}

We combined the datasets (rejecting noisy end channels) into a single
sequence of 79 consecutive frequency channels.  The velocity profile
of the whole spectral data set before continuum subtraction is shown
in Fig.~\ref{VELPROF.PS}, with line emission in channels 19 to 66.
The slanting line shows a linear fit to the continuum-only end
channels.  The slope of this line is probably an artefact due to
interference which was only serious at the low frequency end of the
spectrum, making these data less reliable despite editing. 
The mean
and rms of the continuum bandpass is ($251\pm3$) mJy. We averaged and
imaged the continuum-only channels and used the {\sc clean} components
for self-calibration; the solutions were applied to all data.  After
calibration, the data from each antenna were weighted in proportion to
antenna sensitivity.

To obtain line-only data, we Fourier transformed the visibility data
to make a dirty datacube of images and then subtracted the continuum
using a linear extrapolation between the average of the two groups of
end channels ({\sc aips} task {\sc imlin}). Finally we {\sc clean}ed
the resulting datacube by deconvolving the dirty beam.  We found that
partial uniform weighting (extrapolating into the undersampled parts
of the visibility plane to improve resolution) and using a 120-mas
circular restoring beam gave the best resolution without increasing
the noise noticeably nor losing sensitivity.  This gives an off-source
3$\sigma_{\rm rms}$ noise level of (2-3)~mJy~beam$^{-1}$ per
channel. The noise distribution in the {\sc clean}ed region is
Gaussian and no emission is $<-4\sigma_{\rm rms}$.

We fitted 2-dimensional Gaussian components to measure the position
and intensity of each patch of maser emission in each channel, using
criteria similar to those described in \scite{Yates00}.  The continuum
emission from Mrk 231 is much brighter, and hence the calibration
solutions and continuum subtraction are more accurate than in the case
of Mrk 273.  Moreover, the masers in Mrk 231 cover a larger region and
are easier to separate spatially. We rejected components if they were
$<3\sigma_{\rm rms}$ and did not fall in a series occuring at the same
position (within the errors) in at least 3 consecutive channels at the
start and end of each series. We assumed that the occasional gap
within a faint series was probably due to emission falling just below
our detection threshold in that channel, rather than representing a
significant discontinuity. 

 The errors associated with fitting Gaussian components are analysed
for ideal situations in \scite{Thomson01} and \scite{Condon97};
applications to sparsely sampled arrays are described for VLA
snapshots by \scite{Condon98} and for MERLIN by \scite{Richards99}.
For the present observations the position uncertainty is given by the
naturally weighted beamsize divided by the signal to noise ratio of
each component, plus fitting errors due to non-Gaussian flux density
distribution.  For a 10~mJy~beam$^{-1}$ component the uncertainty is
typically 20~mas.  In many channels the emission was resolved into up
to four components.
\begin{figure*}
\resizebox{18cm}{!}
{\epsfbox{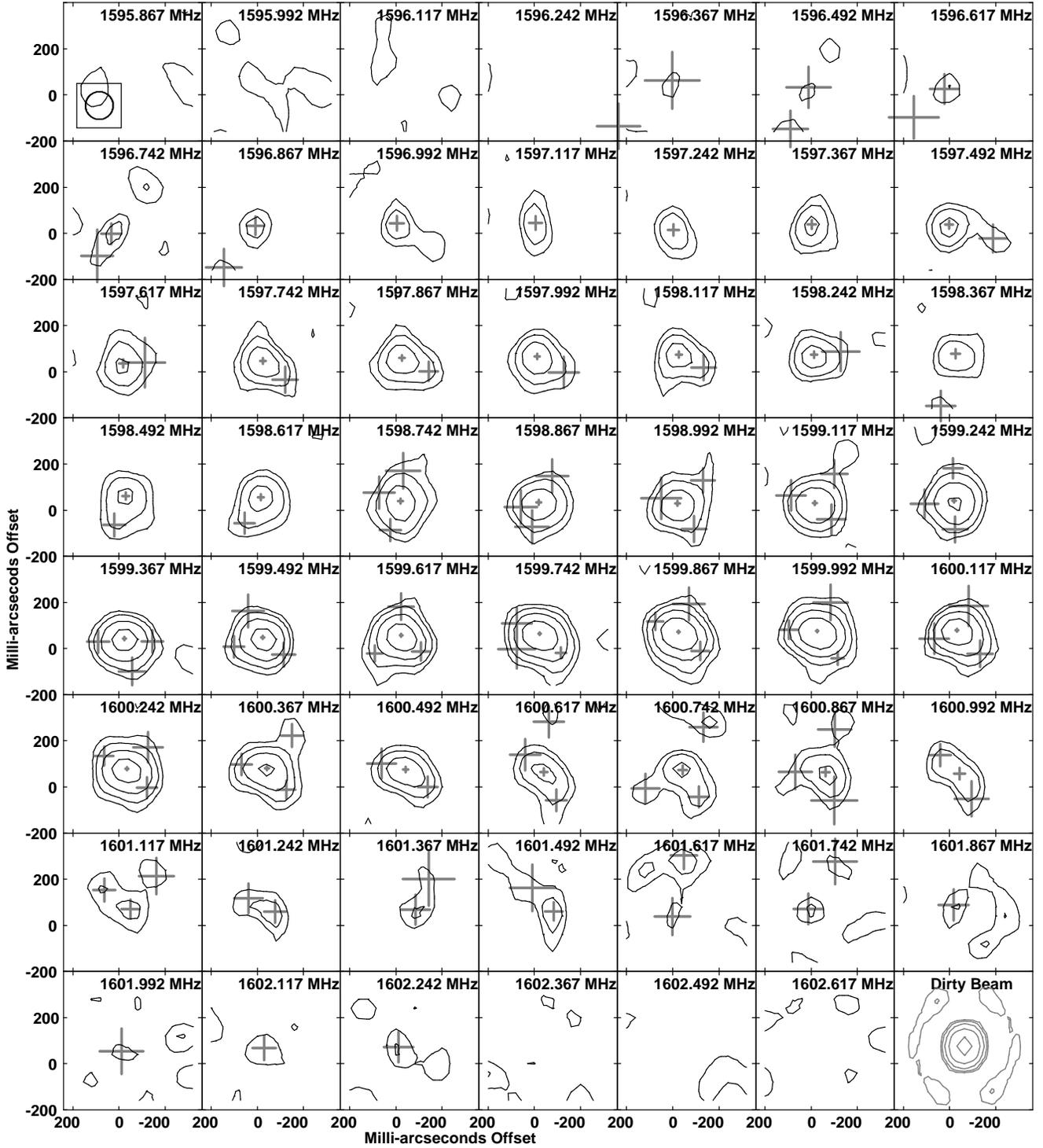}}
\caption{The contours show the OH maser emission in each channel, at
  (1, 2, 4 ...)$\times$ 2 mJy beam$^{-1}$ (the 3 $\sigma_{\rm rms}$
  level in the least noisy channels).  The 120-mas circular restoring
  beam is shown at top left and the dirty beam (contour levels (1, 2,
  4 ...)$\times$ 5 percent of the peak) at bottom right.  The crosses show the error bars for
  fitted maser component positions. The ($x$, $y$) positions are in milli-arcsec offset from the reference position at
  12\h~56\m14\fs~2383 +56\degr~52\arcmin~25\farcs210 (J2000).}
\label{M231_NEW.KPS}
\end{figure*}

\section{RESULTS}
\label{results}

 The results are summarized in Fig.~\ref{M+C.GPS} which shows contours
of the OH emission superimposed on a gray scale image of the radio
continuum.  MERLIN detected continuum emission over 2~arcsec, with a
bright core that is covered by the OH emission. The OH is just over
0.5~arcsec in angular extent, about three times wider than the compact
core detected by \scite{Klockners03} using the EVN.  The OH peak is
slightly displaced (40 mas), from the continuum peak.  The continuum
is weakly polarized in the E--W direction.  The continuum results are
described in more detail in the following Section~\ref{continuum}, and
the OH results in Section~\ref{masers}.

\subsection{Radiocontinuum morphology and polarization}
\label{continuum}
The continuum emission had a similar elongated heart-shaped appearance
in both 1658-MHz images and in the 1420-MHz image. There is no
significant difference in peak intensity (Table~\ref{flux}) nor any
significant position offset or difference in polarization properties
between any of the images so we discuss the 1993 1658-MHz image which
provides the most sensitive highest-resolution data. This is shown in
greyscale in Fig.~\ref{M+C.GPS}.  The continuum peak of
196~mJy~beam$^{-1}$ is located at 12\h~56\m14\fs~2337
+56\degr~52\arcmin~25\farcs237 (total uncertainty 10 mas; uncertainty
relative to the MERLIN maser positions 1 mas). This is (1, --8) mas
from the position reported by \scite{Ulvestad99b} using the VLBA with
phase referencing at 15 GHz and (--6, 0) from the astrometric position
measured by \scite{Patnaik92} using the VLA at 8.4 GHz with an
accuracy of 12 mas.  The continuum emission detected by MERLIN has a
maximum elongation of almost 2 arcsec N--S. Increasing the resolution
by using uniform weighting shows a faint ridge running S from the peak
(corresponding to the trough in the maser:continuum ratio described in
Section~\ref{gain}).
This is in the
direction of the radio jet seen on scales of tens of mas upwards
(\pcite{Carilli98}; \pcite{Taylor99}; \pcite{Ulvestad99sa};
\pcite{Lonsdale03}).  Elongated radio emission is seen on both
larger and smaller scales, see Sections~\ref{intro}, \ref{sb_jet}
and~\ref{misalignments}.

The 1.6-GHz continuum peak brightness temperature is $\ge3\times10^8$
K (derived from \pcite{Lonsdale03}) and all the emission detected by
MERLIN is $>10^5$ K, indicating its synchrotron origin. The total flux
density above 3$\sigma{\rm rms}$, measured from the continuum image
shown in Fig.~\ref{M+C.GPS}, is 265.5(0.4) mJy in 2.49 arcsec$^2$.  In
the continuum peak region, we detected compact linearly polarized
radio emission with a polarized flux density of
$0.78\pm0.07$~mJy~beam$^{-1}$ and a polarization position angle of
$\chi_{\rm rad} = -84\pm2$\degr, shown by the white vectors in
Fig.~\ref{M+C.GPS}. The polarized intensity is $0.4$ per cent of the
total intensity.
There is $\le2$\deg difference between the polarization angles at 1420
and 1658 MHz.  We infer that a core region $\la200$ mas across is
significantly polarized in the presence of a magnetic field oriented
N--S, similar to the direction of the jet on scales of 10--100 pc.

\subsection{Maser distribution}
\label{masers}

Figure~\ref{VELPROF.PS} shows that MERLIN detected OH line flux above
the continuum level across at least 6 MHz of the 8 MHz observing band.
There are two major peaks, one of 53 mJy at 1599.992 MHz and one of 26
mJy at 1597.867 MHz, a separation of 2.125 MHz.  The peaks are within
one spectral channel of the OH mainline frequencies 1667.359 and
1665.402 MHz respectively, redshifted to the velocity of Mrk 231.
Fig.~\ref{VELPROF.PS} shows additional faint emission around 1596 MHz
but this could not be imaged.  This is the region of the spectrum more
affected by interference and it is not possible to tell if the
emission is astrophysical.  The peak flux densities and other details
of the MERLIN spectrum are in good quantitative agreement with the
single-dish spectrum of Stavely-Smith et al. (1987).  The total flux
density of all the fitted components listed in
Table~\ref{M231NEWSPOTS.TAB} is $(1.19\pm0.01)\times10^{-21}$ W
m$^{-2}$, almost exactly the single dish flux density integral of
$(1.12\pm0.12)\times10^{-21}$ W m$^{-2}$ from \scite{Stavely-Smith87}.
We are satisfied that the MERLIN images represent the main features of
the maser distribution accurately.

Fig.~\ref{FreqDec.GPS} shows a declination-frequency plot of the OH
emission above 6 per cent of the peak (about $3\sigma_{\rm rms}$). The
brightest emission, between 1599 and 1602 MHz, shows an increasing
displacement towards higher declination with higher frequency.  A
fainter copy of this pattern is seen between 1596.5 and 1598.5 MHz.
The 1667-MHz line is almost always brighter than the 1665-MHz line in
megamaser galaxies \cite{Henkel90}, by a factor similar to the LTE (local
thermodynamic equilibrium) line ratio of 1.8:1, or greater.  We
therefore assume that the brighter emission is dominated by the
1667-MHz line and the fainter copy by the 1665-MHz line.  There is
emission across the whole frequency range near the centre of the
declination range. There is also a halo of fainter emission.

 The contours in Fig.~\ref{M+C.GPS} show all maser emission summed in
frequency; the maser peak at 12\h~56\m14\fs2351
+56\degr~52\arcmin~25\farcs274) is significantly offset by (12, --37)
mas (uncertainty 2 mas) from the continuum peak.  The total maser flux
density in fitted components in the channels corresponding to the
1667- and 1665-MHz peaks is 48 and 21 mJy~beam$^{-1}$ respectively. In
both cases there is a difference of $5\pm4$ mJy between the spectral
flux (Fig.~\ref{VELPROF.PS}) and that appearing in fitted components,
suggesting that an insignificant amount of maser emission has escaped
fitting.

Channel maps of the maser emission with the positions of the fitted
components overlaid are given in Fig.~\ref{M231_NEW.KPS}. Negative
contours are not shown in the crowded plots but there is no negative
emission $<-4\sigma_{\rm rms}$. The dirty beam pattern is shown in the
last panel.  

We allocated the fitted components to one or other
transition using these assumptions:
\begin{itemize}
\item{The brightest component in the main/secondary peak originated at
  1667/1665 MHz;}
\item{The most extreme blue-/red-shifted components originated at
  1667/1665 MHz;}
\item{Components with a similar angular position separated by $\approx2$ MHz in
  frequency arose from both transitions.}
\end{itemize}
 
\begin{figure}
\resizebox{8.5cm}{!}{\rotatebox{-90}
{\epsfbox{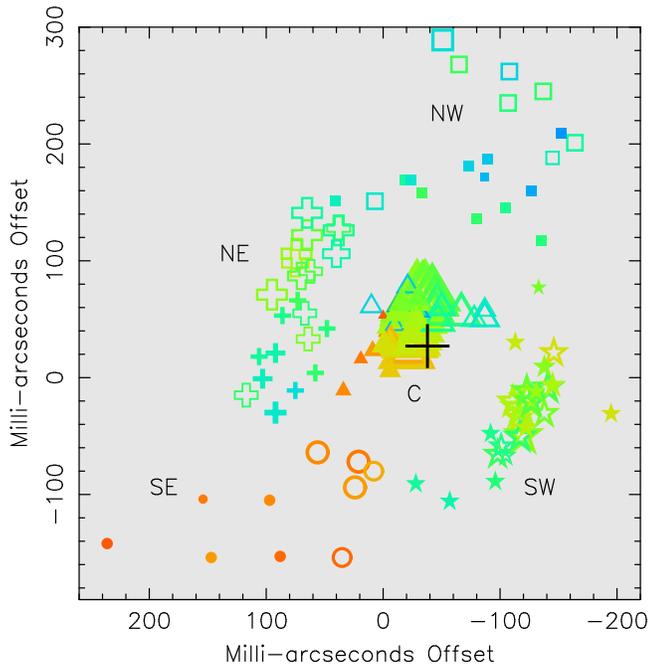}}}
\caption{OH 1665- and 1667-MHz maser components in Mrk 231 are shown
  by solid and hollow symbols respectively.  The different symbol
  shapes identify the five emission regions (NE, SE, SW, NW and C)
  described in Section~\ref{masers}.  The symbol area is proportional
  to the component flux density (the brightest central components are
  slightly reduced in size for clarity). Position errors  (not
  shown) range from 5--100 mas and are inversely proportional to flux
  density. The exact parameters are given in
  Table~\ref{M231NEWSPOTS.TAB}. The cross marks the position of the
  peak 1.6 GHz continuum emission, with an uncertainty of 1 mas. ($x$,
  $y$) positions are relative to the reference position at
  12\h~56\m14\fs~2383 +56\degr~52\arcmin~25\farcs210 (J2000).}
\label{M231NEWSPOTS.CPS}
\end{figure}

Seventy-one components were allocated to five spatially separate
 1667-MHz features and 54 components were allocated to corresponding
 features in the 1665-MHz line. These regions are labelled on
 Fig.~\ref{M231NEWSPOTS.CPS}.  The brightest region, C, occurs near
 the centre of the maser distribution. The peak flux density of 38 mJy
 beam$^{-1}$ occurs in the 1667-MHz transition at $V_{\rm LSR}$ 12113
 km s$^{-1}$ ($V_{\rm hel}$ 12613 km s$^{-1}$), position ($x$, $y$) =
 (--29, 66) mas, $\sigma_{xy}=5$ mas.  The closest, brightest emission
 in the 1665-MHz transition, 17 mJy beam$^{-1}$, occurs in the
 adjacent channel, $V_{\rm LSR}$ 12135 km s$^{-1}$ ($V_{\rm hel}$
 12636 km s$^{-1}$), at (--14, 56) mas, $\sigma_{xy}=5$ mas.  Emission
 at the same intensity is co-located within the position errors, two
 channels away in velocity at $V_{\rm LSR}$ 12180 km s$^{-1}$ ($V_{\rm
 hel}$ 12685 km s$^{-1}$),.  The combined error-weighted average maser
 peak position is (--26, 64) mas, uncertainty 5 mas.  Additional
 blue-shifted emission occurs in the region labelled NW and more
 red-shifted emission in SE.  The regions marked NE and SW contain
 emission at intermediate velocities.  

The core region C contains most of the total maser flux, 71 per cent
 whilst the halo regions NE, SE, SW and NW contain 9, 3, 11 and 6 per
 cent of the total, respectively.  The outer regions have peaks of 30
 per cent or less of the region C peak for the same transition.

 The peaks are well-separated but the edges of some or all of the five
regions could be overlapping in space and/or velocity and some
component blending is likely. For example, 1667-MHz emission from the
brightest maser region (labelled C) appears spectrally asymmetric with
a longer blue-shifted wing whilst the opposite is true at 1665
MHz. This is probably due to truncation of the lines where it is
difficult to separate broad components in regions where they overlap
in position and velocity.  If each line has an intrinsically symmetric
velocity profile then components in 7--8 channels ($\approx150$
km~s$^{-1}$) are missing from the fits to the overlap region.  This
may also be the case to a lesser extent for the NE, SW and NW regions,
in addition to emission which is below our sensitivity level. It is
unlikely that more than one or two components at each overlap have
been wrongly allocated (i.e. less than 10 per cent of all components).

In all regions there is a tendency for the 1665-MHz
emission to be offset to the SE from the 1667-MHz emission. This could
be due to misalignment of the two input data sets due to uncertainties
in phase-referencing but this is very unlikely to exceed a few mas due
to the use of a very nearby phase reference source
(Section~\ref{lineobs}) and the fact that the observations were
interleaved. The offset is often less than the position uncertainty
due to noise for individual components so we cannot speculate about
any possible astrophysical origins.

 We used the rest frequencies to estimate the Doppler velocity
 ($V_{\rm LSR}$) of the emitting gas for each component.  In this and
 similar contexts red- and blue-shifted are used to describe
 velocities with respect to the systemic velocity of Mrk 231.  The
 component parameters are tabulated in
 Table~\ref{M231NEWSPOTS.TAB}. Column (1) gives the region
 identification, column (2) gives the channel observing frequency and
 columns (3) and (4) give the velocities  in the optical heliocentric
 and radio LSR conventions.
 Columns (5)--(7) give the component offsets ($x$, $y$) from the
 reference position and the position uncertainty $\sigma_{xy}$ and
 column (8) gives the component flux density $P$ with an uncertainty
 of $\le1$ mJy beam$^{-1}$.

\section{The structure and dynamics of Mrk 231}
\label{dynamic}

\subsection{A compact, polarized core}
\label{polarization}

The radio emission within $\le80$ pc of the core is $\approx0.4$ per
cent polarized with $\chi_{\rm rad} = -84\pm2$\degr, which implies a
N--S magnetic field (Section~\ref{continuum}). \scite{Ulvestad99sa}
report a similar core polarization angle on the VLA (2.25-arcsec)
scale, but the fractional polarized intensity is $<0.1$ per cent,
suggesting that beam dilution reduces the VLA measurement.
\scite{Smith04} measured $\chi_{\rm opt}=(95\pm0.4)$\deg at 5000--7000
\AA\/ using the William Herschel telescope, as part of a survey of
polarized optical emission from Seyfert galaxies. They place Mrk 231
in a class of galaxies showing scattering of AGN light in regions
between 1--300 pc from the nucleus along a polar axis at an angle of
inclination $\approx45$\degr, with the direction of outflow
perpendicular to the polarization angle. Its polarization properties
are similar in the NIR \scite{Jones89}.  {\em ISO} polarimetry by
\scite{Siebenmorgen01} at 12.0 and 14.3 $\mu$m gave $\chi_{\rm
MIR}\approx125$\degr. The authors state that this is in relative
agreement with the NIR results although the MIR polarization is
probably not due to scattering but to dichroic spheroidal dust grains
aligned with the magnetic field in a torus perpendicular to the
outflow. If there is a genuine difference between $\chi_{\rm MIR}$ and
measurements at other wavelengths, it is in the same direction as the
difference in position angle between the pc-scale jet
\cite{Ulvestad99b} and its direction on larger scales (see
Section~\ref{misalignments}).  This implies that the MIR emission
originates from the innermost torus.

 \scite{Soifer00} took Keck images of Mrk 231 at 7 wavelengths between
 7.9--17.9 $\mu$m at a resolution similar to MERLIN at 1.6 GHz and
 retrieved longer-wavelength IRAS data. They deduce that at
 $\lambda\le12.5$ $\mu$m, the IR source is $\le130$ mas in diameter,
 increasing to 400 mas at 60 $\mu$m (assuming a dust temperature of 85
 K in the extended region).  This shows that either there is a
 temperature gradient (decreasing with distance from the core), or
 that the shorter-wavelength emission is dominated by AGN-related
 activity, whilst a cooler region which is a few hundred pc in extent
 is significant at longer IR wavelengths. The latter explanation is
 consistent with the association of polarized radio and
 shorter-wavelength IR emission with the compact base of the jet and
 with emission of starburst origin around 60-$\mu$m dominating in the
 masing region.  The polarization mechanisms differ in the various
 wavelength regimes but the similarities in polarization angle within
 the inner few hundred pc support our assumption that $\chi_{\rm rad}$
 is the same at 1420 and 1658 MHz because Faraday rotation is
 negligible (rather than some exact multiple of $\pi$).

\subsection{Maser kinematics}
\label{kinematics}
\subsubsection{Sub-kpc scales}
\label{sub-kpc}
Regions SW and NE (marked in Fig.~\ref{M231NEWSPOTS.CPS}), separated
by ($360\pm50$) mas, define the minor axis of an ellipse containing
the masers and cover the central 400 km s$^{-1}$ of the velocity
range. Regions SE and NW contain only red- and blue-shifted emission
respectively, including the most extreme velocities (separated by 720
km s$^{-1}$) and have an angular separation of ($510\pm70$) mas which
defines the maser major axis $B_{\rm maj}$.  The $V_{\rm LSR}$
gradient between SE and NW is $dV_{\rm LSR}/dB_{\rm maj} =
(1.4\pm0.2$) km~s$^{-1}$ mas$^{-1}$ $\approx1.7$ km~s$^{-1}$
pc$^{-1}$.  This suggests rotation (towards the observer in the NW) at
radii $r \le (210\pm30)$ pc.  The axis of rotation, projected
against SW and NE, is at a position angle of $230$\degr$\pm10$\degr.
The axial ratio of the ellipse containing the masers shows that the
rotation axis is at an angle of inclination to the plane of the sky of
$i = 45$\degr$\pm10$\degr.  The radio jets appear to point
approximately S on all but the smallest scales (Section
\ref{continuum}) 
but the maser axis direction is less inconsistent with the jets if it
is pointing towards the observer in the SE.  We will refer to the
masing region as a torus although it could have some other shape such
as a disc.

The smallest velocity gradient along the line of sight is found
projected against the rotation axis in a thin torus in Keplerian or
solid body rotation.  The largest velocity gradient along the line of
sight would then be $\la1.7$ km~s$^{-1}$ pc$^{-1}$ for regions in the
orthogonal direction in the plane of the sky. The lower velocity
gradient may be allowing greater maser amplification (in our
direction), since regions NE and SW contain $\ge2\times$ the flux of
NW and SE.  

The total velocity gradient perpendicular to the rotation axis is
given by $dV/dr = (dV_{\rm LSR}/dB_{\rm maj})/\sin{i} = (2.4\pm0.4)$
km~s$^{-1}$ pc$^{-1}$ where $V$ is the maser rotation velocity in
km~s$^{-1}$. The enclosed mass density $\rho_{\rm M}$  is given by
\begin{equation}
\rho_{\rm M} = \frac{(dV/dr)^2}{(4/3)\pi \times 0.0043}
\label{encmass}
\end{equation}
in units of M$_{\odot}$ pc$^{-3}$.  This gives $\rho_{\rm M} =
320\pm90$ M$_{\odot}$ pc$^{-3}$ and a maximum enclosed mass of
$\sim10^{10}$ M$_{\odot}$, if it has a uniform spherical distribution
or is a compact object.  If the mass distribution is flattened, or the
torus is nearer edge-on, this would reduce the enclosed mass by a
factor of up to about two.  \scite{Davies04} deduce from 
NIR data that (depending on the star formation history) a total
mass of stars of at least $1.6\times10^{9}$ M$_{\odot}$ is contained
within the inner 200 pc in a disc-like distribution. This would amount
to $\sim1/3$ of the dynamical mass implied by our OH data, assuming a
flattened distribution.

The EVN observations by \scite{Klockners03} detected just over half
the OH maser emission imaged by MERLIN, covering a 200 km s$^{-1}$
velocity range within the innermost 150 mas, a similar angular extent
to the MERLIN region C.  The EVN did not detect the fainter red- and
blue-shifted tails seen in MERLIN regions C, SE and NW, nor most of
the extended intermediate-velocity regions NE and SW.
\scite{Klockners03} measure a velocity gradient corresponding to 1.2
km~s$^{-1}$ mas$^{-1}$, similar to the value of $dV_{\rm LSR}/dB_{\rm
maj}$ which we measure over 510 mas.  They present a model of a torus
inclined at 56\deg rotating about an axis at position angle 35\deg
(their fig.~3{\bf c}), almost parallel to the axis of symmetry deduced
from the MERLIN data. They deduce a central mass concentration of
$(7.2\pm3.8)\times10^7$ M$_{\odot}$ within a radius of $\approx60$
pc. This is consistent with the mass concentration measured using
MERLIN if the mass distribution between 50--200 pc is flattened.  The
inner radius of the torus $r_{\rm i}$ is unresolved by the EVN, so
$r_{\rm i}\le 20$ mas and continuing the extrapolation would suggest
that $\la5\times10^6$ M$_{\odot}$ is enclosed.

\subsubsection{Innermost masers}
\label{inner}
The distinct `core-halo' maser distribution seen in
Fig.~\ref{M231NEWSPOTS.CPS} suggests that this is not the whole story.
If the kinematics of a torus are determined by a smooth mass
distribution, or by a central point-like mass alone, we would expect
the maser emission to have a smooth radial profile in any given
direction. For solid-body rotation, the most extreme velocities should
be seen at the outer rim.  This is a reasonable description of the
broad ring traced by NE, NW, SW and SE.  The exponential nature of
unsaturated maser amplification means that small differences in local
density or turbulence can be greatly exaggerated so small fluctuations
should not be over-interpreted.  However, the flux density gradient
steepens towards the central region C (which is not spatially resolved
by MERLIN); this is best seen in the linearly-spaced contours of
Fig.~\ref{FreqDec.GPS}.  Region C includes weak tails covering the
total $V_{\rm LSR}$ range of 720 km s$^{-1}$ within 120 mas (100 pc).
Interaction between the jet and the torus is a possible but unlikely
explanation, see Section~\ref{misalignments}.  A higher mass
concentration is more probable. This could imply a very dense nuclear
starburst but there is no evidence for this in results of
\scite{Davies04s} and \scite{Soifer00}.

We suggest that the extreme velocity gradient is due to a supermassive
black hole starting to dominate the dynamics at $r_{\rm i}$. Applying
Equation~(\ref{encmass}) gives an enclosed mass of $\sim8\times10^6$
M$_{\odot}$ within $r_{\rm i}\la16$ pc (for $i\sim45\degr$).
\scite{Ferrarese02} did not consider any black hole mass estimates
below $3\times10^7$ M$_{\odot}$ secure, apart from those in the Milky
Way and M32.  The mass of the nucleus of Mrk 231 was estimated at
$1.3\times10^8$ M$_{\odot}$ by \scite{Padovani90}. This was based on
the velocity width of the H$\beta$ line and an estimate of the inner
radius of the broad-line region (BLR) of $\approx0.01$ pc, derived
from the UV:X-ray spectral index, but, as noted by
\scite{Padovani90s}, the relationship used is only reliable in a
statistical sense. \scite{Ferrarese02} found that it could be in error
by an order of magnitude in comparison with less model-dependent
reverberation mapping.  More recent {\em Chandra} observations by
\scite{Gallagher02} showed that X-ray emission from the unresolved
nucleus of Mrk 231 is variable on a timescale of hours, suggesting an
origin within tens of AU.  The 2 keV flux is underluminous with
respect to the UV continuum and is best fitted by a model of
absorption and scattering on scales $\la0.001$ pc.  If this is so,
then the black hole mass estimated using the BLR radius is reduced by
at least an order of magnitude, closer to our value.

\subsection{Maser clouds}
\label{clouds}

The component positions (fitted as explained in Section~\ref{lineobs})
measure the locations of the maximum line-of-sight amplification of
maser emission from unresolved clouds within the area enclosed by the
contours in Fig.~\ref{M+C.GPS}.  We use the term cloud to represent a
discrete region which may be defined by the density, temperature or OH
abundance required for masing, or by the local velocity gradient.  The
individual maser components are unresolved so that the size of the
restoring beam ($\approx100$~pc) provides the highest upper limit to
cloud size.

OH amplification occurs over at least the distance wherein the
velocity changes by less than the thermal line width, which is
$\sim0.4$ km~s$^{-1}$ at the probable temperature of 85 K in the maser
region \cite{Soifer00}.  The model of a rotating torus described in
Section~\ref{kinematics} leads to an estimate of $\approx1.7$
km~s$^{-1}$ pc$^{-1}$ for the maximum line of sight velocity gradient,
implying a minimum gain length of 0.25 pc. \scite{Randell95} show that
propagation is possible over a larger velocity difference, see
Section~\ref{physics}.  

The OH masers were resolved out by \scite{Lonsdale03} using global
VLBI at 5-mas (4-pc) resolution but, using a tapered beam (coarsened
resolution), they detected weak emission from the 1667-MHz line
only. The position and peak flux density appear similar to those in the EVN
image.
This implies a maximum cloud size  $<4$ pc, since maser
emission from a single 5-mas cloud would be beamed into a much smaller
angle, but the  emission per interferometer beam
from a collection of clouds of smaller radii decreases in
proportion to the beam area.  
These constraints
 suggest that the masers
propagate through clouds in the size range of 0.25--4 pc (if they are
approximately spherical) giving a typical diameter $l\sim1$ pc.

\scite{Ulvestad99sa} infer an ionised number density of $\sim10^9$
m$^{-3}$ at 20 pc from the nucleus (see Section~\ref{misalignments}),
close to our estimate of $r_{\rm i}$. This is only one or two orders
of magnitude smaller than the total number density supporting OH maser
emission (see \pcite{Randell95} as applied in Section~\ref{physics})
and such a high ionised fraction seems unlikely in clouds cool enough
to support masing, suggesting that the torus contains cool dense
molecular clumps embedded in a highly ionised
medium. \scite{Carilli98s} present a disc model in which the inner 20
pc is predominantly ionised, the next 100 pc contains neutral atomic
gas and molecules and dust extends out to 400 pc, with a height:depth ratio
$<3:40$.  Our results broadly support this, as \scite{Carilli98s} also
note that the phases may be present in mixtures of differing
proportions at different radii, allowing the presence of OH within 120
pc of the nucleus.  The thickness of the OH torus would then be $\sim$
10 pc or less. A similar stratification of all 3 phases in a thin disc
has only been directly measured in a few other active galaxies such as
NGC 4261 (\pcite{Jaffe96}; \pcite{Jones97}; \pcite{vanLangevelde00}).
This is not, however, the whole story, as the disc is also warped, see
Section~\ref{misalignments}.

\subsection{Starburst and jet emission}
\label{sb_jet}
\begin{figure}
\resizebox{8.7cm}{!}
{\epsfbox{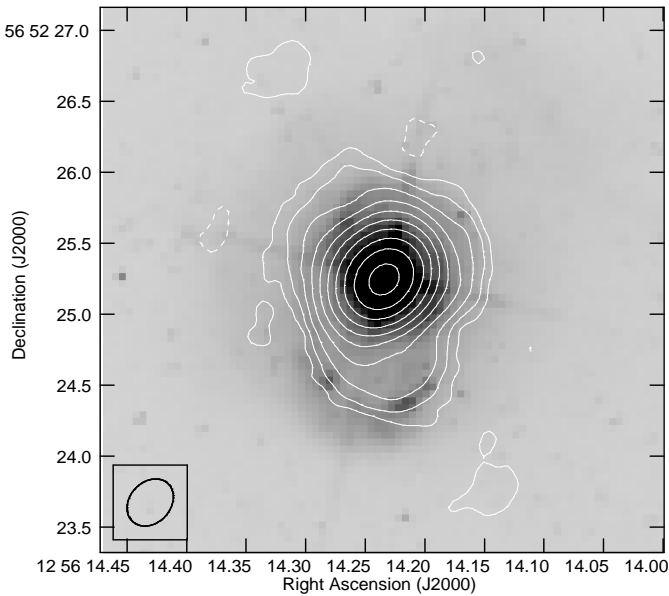}} 
\caption{The greyscale shows an {\em HST} WFPC2 image taken using the
  F814W (red) filter at a pixel resolution of 46 mas. The contours
  show the 1.6-GHz radio continuum (also shown in grey scale in
  Fig.~\protect{\ref{M+C.GPS}}) at levels of (--1, 1, 2, 4...)$\times0.3$ mJy beam$^{-1}$;
  the restoring beam is shown at lower left. }
\label{HST+MERLINM231.PS}
\end{figure}

\scite{Ulvestad99sa} and \scite{Taylor99} deduce that the radio
continuum emission within $\sim1$ kpc of the nucleus is mostly of
starburst origin, probably triggered by a merger $10^8$--$10^9$ years
ago.  A {\em Hubble Space Telescope} Wide Field and Planetary Camera 2
({\em HST} WFPC2) image taken using the F814W (red) filter
(Fig.~\ref{HST+MERLINM231.PS}) shows bright knots which are probably
sites of star formation, near the southern rim of the MERLIN radio
contours (corresponding to a 1.6-GHz brightness temperature of
$\approx5\times10^5$ K).  The transfer function for the optical image has
been chosen to show extended emission at the expense of saturating the
brightest regions.  There is in fact a compact core and the images
were registered by aligning this with the radio peak.  The absence of
strong optical emission at intermediate distances does not rule out a
starburst origin for the radio emission since obscuring dust would
only allow us to detect the longer wavelengths.  The core could still
be visible through a central cavity in the dusty disc as suggested
by the free-free absorption model of \scite{Taylor99}.

Whilst star formation can explain most of the smooth extended emission
on MERLIN scales, the jet and its base are likely to be the source of
the southern ridge and the polarized emission, supporting an N--S
direction of the approaching jet from a few tens to a few hundred pc
from the nucleus. This seems to be in front of the OH torus, see
Section~\ref{gain}.  \scite{Taylor99} show that the spectral index in
this direction is steeper than in the surrounding diffuse emission,
suggesting a jet extending due south out to $\sim180$ pc. This is
supported by other VLBA and global VLBI images with 10 -- 100 mas
resolution (\pcite{Carilli98s}; \pcite{Lonsdale03}). 
VLA observations on larger scales show a 50-kpc region of diffuse
1.4-GHz emission extended to the south (\pcite{Carilli98s};
\pcite{Ulvestad99sa}).  This is not aligned with the apparently tidal
optical tails seen on a similar scale to the north and south
\cite{Hutchings87}. WSRT observations reveal radio emission out to 120
kpc \cite{Baum93}. The extended radio emission may be associated with
star formation (e.g. as a result of a superwind) and/or with the
extension of the jet (see discussion in \pcite{Ulvestad99sa}). 

\subsection{Misalignments}
\label{misalignments}

 The axes of symmetry of the OH velocity distribution deduced from
 observations using MERLIN (Section~\ref{kinematics}) and the EVN
 \cite{Klockners03} are within 10\deg of a NW-SE direction.
 \scite{Carilli98s} used the VLBA to measure HI absorption modelled by
 a torus with an axis at an angle of inclination of 45\degr, similar
 to that of the OH torus.  Absorption is not detected towards the core
 nor the brighter jet lobe, reinforcing the model in which the
 southern/western jet is approaching and the neutral material is in a
 torus with a sufficiently large central cavity ($r_{\rm i} > 5$ mas)
 that the core is not obscured at our viewing angle.  However, the HI
 velocity gradient, of 1.15 km~s$^{-1}$ mas$^{-1}$ across 200 mas, runs W
 to E, giving an N--S symmetry axis. The magnitude of the velocity
 gradient is consistent with solid body rotation of a flattened mass
 distribution as discussed in Section~\ref{kinematics}.  CO emission
 on arcsec scales \cite{Bryant96} suggests an almost face-on disc
 rotating about an axis running N--S.  These results imply
 misalignements of $\approx45$\deg in both spatial axes between the OH
 and CO emission regions, with the HI disc partially aligned with
 both. These misalignments can be understood in terms of a warped disc
 seen almost face on.  The kinematic axis of such a model can shift
 with distance from the centre and even appear to change direction
 (see fig.~3 in \pcite{Cohen79}). A warped disc was also found in the
 inner 0.3 arcsec of Mrk 231 by \scite{Davies04s} using Keck NIR $H$-
 and $K$-band spectroscopy of stellar absorption, at an angle of
 inclination of only 10\degr.  Alternatively, misaligned orbital
 systems of ionised and neutral gas are seen in e.g. M82, explained by
 \scite{Wills00} using the barred potential model of
 \scite{Binney87}. Such models cannot, however, explain fully the
 segregation of different neutral species seen in Mrk 231.

The highest resolution images \cite{Ulvestad99b}, made using the VLBA
at 15 GHz, show a mas-scale jet at a position angle of 245\deg.  All
other available radio images show that from $\sim15$ mas outwards the
jet (and counter jet, where detected) run approximately N-S
(Section~\ref{sb_jet}).  The OH axis on scales of 60--250 mas is at an
angle close to the direction of the jet in the innermost few mas, so
we must examine whether the jet is being deflected by collision with
the masing torus.  If this produced a large shocked slab, efficient
maser amplification could occur perpendicular to the shock
\cite{Elitzur92}.  If the mas-scale jet was at $\la1$\deg to the line
of sight, its deprojected trajectory would be comparable to the size
of the maser region.  Proper motion measurements imply that a jet at
such a small angle would be strongly relativistic, but
\scite{Ulvestad99b} measure an apparent proper motion of only $0.14c$.
They argue that free-free absorption is a better explanation (than
relativistic boosting) for the high jet-counterjet brightness ratio.
Moreover, in order to impede the jet, the OH torus should be close to
edge-on, but in fact it is at an angle of inclination of $\sim45$\degr,
making the jet collision hypothesis very unlikely on the scale of the
maser region.

 \scite{Klockners03} found a region of high velocity dispersion in the
 north west of the inner part of the torus which they suggest could be
 due to jet-cloud interactions.  We found even higher anomalous
 velocities in the central region but in Section~\ref{inner} we
 showed that could be due to the black hole. As \scite{Carilli98s}
 also point out, outflows would produce a separation between red-and
 blue-shifted molecular emission, but the detailed velocity gradients
 resolved by HI and OH observations are more characteristic of
 rotation than of outflow although local anomalies caused by
 interactions are possible.

\section{Maser amplification constraints}
\label{maser}

\begin{figure}
\resizebox{8.5cm}{!}
{\epsfbox{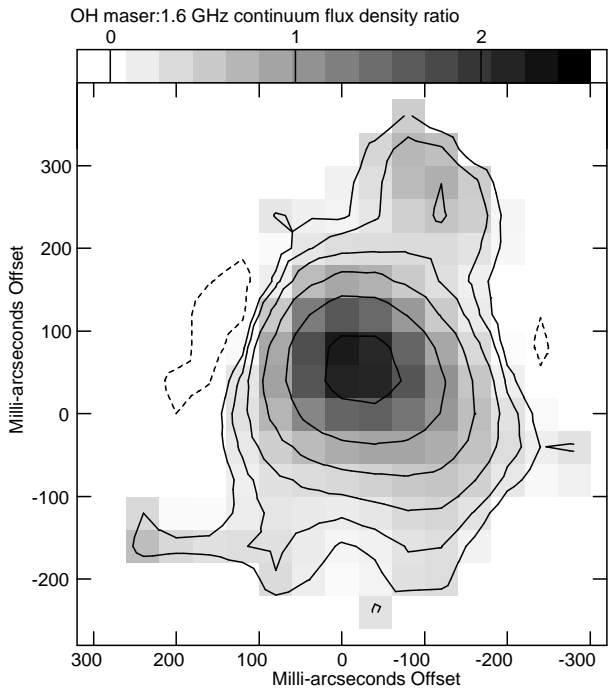}} 
\caption{The greyscale shows the ratio of the total OH maser emission
  to the continuum flux at 1658 MHz at the same resolution (120 mas
  beam).  The contours show the sum of all OH maser emission as in
  Fig.~\ref{M+C.GPS}.  }
\label{M231DIVOH.GPS}
\end{figure}

\subsection{Maser gain}
\label{gain}

All the maser emission is projected against a region enclosed by the
20 mJy~beam$^{-1}$ contour of continuum emission at ($362 \times 284$)
mas$^2$ resolution. The corresponding 1.6-GHz continuum brightness
temperature of $\ga10^5$ K is a more effective source of 1.6-GHz seed
photons than spontaneous emission from OH at 85 K.  Summing all the
maser emission from both transitions into one image gives a peak at
 ($x$, $y$) = (--28, 49) mas, 22
mas from the continuum peak (Section~\ref{continuum}).  We use the
continuum image at 120-mas resolution for comparison with the summed
maser-only image in order to estimate the ratio $R_{\rm M/C}$ of maser
emission to 1.6-GHz continuum emission, shown in
Fig.~\ref{M231DIVOH.GPS}.  The lowest maser contour is error-dominated
and not used in analysis. The maximum $R_{\rm M/C}\approx2.2\pm0.1$
occurs in the brightest central region, at ($x$, $y$) = (--22, 55) mas with an
uncertainty $\sigma_{xy}$ 7 mas, located 8 mas from the peak of the
combined maser emission. Note that the apparent $R_{\rm M/C}$ will be
an underestimate in any regions where appreciable continuum emission
is in the foreground with respect to the masers.  The site of the
maximum ratio is a more significant distance of ($24\pm10$) mas north
of the continuum peak, in the direction of stronger HI absorption
\cite{Carilli98s}, suggesting that, in this region, the observed
neutral material is in front of the radio continuum and that the
maximum observed gain is close to the true value.  In the whole area
with significant maser emission the average $R_{\rm M/C}\approx0.8$.

 Fig.~\ref{M231DIVOH.GPS} shows that $R_{\rm M/C}$ is higher in the NW
region than in SE and SW and there is a slightly lower-ratio trough
running due S from the centre, in the direction of the jet on similar
scales (see references in Section~\ref{sb_jet}). Bright
continuum hotspots which reach $2\times 10^9$ K at 1.6 GHz are
revealed by global VLBI \cite{Lonsdale03} but these are not coincident
with enhanced OH emission.  The brightest radio continuum might
originate in regions hostile to masers and/or in the foreground, such
as from the approaching jet.  Both would lead to an underestimate of
the maser gain.


The peak of the integrated OH emission imaged by the EVN observations
of \scite{Klockners03} is $\approx5$ mJy beam$^{-1}$.  This is 1/13 of
the  MERLIN peak, the same ratio as that between the
EVN and MERLIN beam areas.  Maser emission outside the central 150
mas would have a maximum flux density of $\sim1$ mJy in the EVN beam,
just below the limit for reliable detection. \scite{Klockners03}
detect continuum emission down to 4 mJy beam$^{-1}$ in the innermost
70 mas of their maser region but the brightest maser emission is
outside any detectable continuum at their resolution.  They argue that
this indicates a maser gain of $\approx2.2$ in all but a few patches,
consistent with the MERLIN value of $R_{\rm M/C}$ for region C.  This
is confirmed by scaling the MERLIN  continuum flux density to the
EVN resolution, which shows that all the maser emission detected by
the EVN is associated with radio continuum $>0.8$ mJy per EVN beam.
The EVN maser -- continuum peak offset is similar to the offset seen
in the MERLIN images.

The hyperfine ratio $R_{\rm H}$ between emission from the 1667- and
1665-MHz transitions is often quoted as the ratio of the peaks in the
frequency spectrum, $2.3\pm0.4$ in Fig.~\ref{VELPROF.PS}, but this
could be misleading as both peaks contain tails of emission from the
other transition.  Comparing the total resolved emission from the
images could lead to an overestimate as a greater proportion of the
fainter 1665-MHz transition is likely to be below our detection
limit. To get the most accurate hyperfine ratio we compared the peak
flux densities in each of the 5 spatial regions.  The error in each
case (derived from the signal to noise ratio) is $\approx0.2$.  The
maximum value of $R_{\rm H}=2.2$ occurs in the brightest region,
C. The ratios in the other 4 regions have a flux-weighted mean of
$1.5$. The overall mean is $1.8$, the same as the LTE line ratio,
indicating that the masers are unsaturated and that the opacity is low
at the masing frequencies. In this regime, the relationships expressed
by \scite{Henkel90} are applicable and we use them to derive the
filling factor.  The 1667:1665 MHz line ratio is related to the
unsaturated optical depth of the 1667-MHz line $\tau_{\rm u}$ by
\begin{equation}
R_{\rm H} = \frac{e^{-\tau_{\rm u}}-1}{e^{-\tau_{\rm u}/1.8}-1}
\label{hyperratio}
\end{equation}
The line-to-continuum ratio $R_{\rm M/C}$ is derived from the images
containing emission from both lines, so the ratio for the 1667-MHz
line alone is given by $R_{\rm M/C}[R_{\rm H}/(R_{\rm H}+1)]$.  The
apparent optical depth $\tau_{\rm a}$ at 1667 MHz is derived from the
line-to-continuum ratio using
\begin{equation}
\tau_{\rm a}=-\ln\{1+R_{\rm M/C}[R_{\rm H}/(R_{\rm H}+1)]\}
\label{apptau}
\end{equation}
and this is related to the maser cloud covering factor $F$ by
\begin{equation}
F=\frac{e^{-\tau_{\rm a}}-1}{{e^{-\tau_{\rm u}}-1}}
\label{filling}
\end{equation}

Using Equation~(\ref{apptau}), we obtain $\tau_{\rm a}=-0.9$, $-0.3$ and
$-0.4$ for the central peak, the other regions and the overall
average, respectively.  The error in the central region is 0.1;
elsewhere the dispersion of $\approx0.3$ dominates the uncertainty.
For each region, if we take $\tau_{\rm u}\approx\tau_{\rm a}$,  
Equation~(\ref{hyperratio}) is satisfied  within the uncertainties,
 so from Equation~(\ref{filling}) we deduce that $F\approx1$.  
Note that \scite{Henkel90} obtained values of $|\tau_{\rm u}|=0.98$,
similar to our peak value and $|\tau_{\rm a}|=0.27$, similar to our
overall average.  This is likely to be due to the good spectral
resolution but low angular resolution of the data then available and
comparison of such values leads to an underestimate of the covering
factor if inhomogeneities in the masing region are unresolved.

 A covering factor close to unity is supported by the lack of any very
bright compact masers (\scite{Lonsdale03}; Section~\ref{clouds}) since
two or more clouds overlapping along the line of sight produce
spectacular maser amplification (e.g.  \pcite{Kartje99}) which favours
1667-MHz emission ( \pcite{Randell95}; see
Section~\ref{comparisons}). This is not seen: higher resolution VLBI
observations (including upper limits) by \scite{Klockners03} and
\scite{Lonsdale03} (see Section~\ref{clouds}) are consistent with a
similar hyperfine ratio on all angular scales in Mrk 231.

\subsection{Physical properties of masing regions}
\label{physics}

 The megamaser model of \scite{Randell95} provides constraints on the
physical conditions producing the Mrk 231 masers.  Their `standard'
model shows that a velocity change of 1.7 km s$^{-1}$ across a 3-pc
cloud supports mainline maser gains in the range we observe.  It is
also compatible with the 1667:1665-MHz line ratio $R_{\rm H}$ which we
observe and with the relative strengths of the satellite lines
(\pcite{Baans92}; see Section~\ref{intro}). \scite{Randell95} do not
consider clouds less than 3 pc in size but find that emission is
suppressed in substantially larger clouds and that smaller velocity
gradients produce a much higher value of $R_{\rm H}$ than is seen in
Mrk 231.

In Section~\ref{clouds} we estimated a typical maser cloud size $l$ of
1 pc, over which the velocity change  $\Delta_{\rm V}$ = 1.7 km
s$^{-1}$.  The size is uncertain to within at least a factor of 4 but
within a regular, thin, torus we can assume that the velocity gradient
$l/\Delta_{\rm V}$ is constant along any particular line of sight. 
 Using equation 9.1.2 from
\scite{Elitzur92},  in our units,
\begin{equation}
\tau_{\rm u} = 14\times10^{-3}\frac{n_{\rm OH}l/\Delta_{\rm
    V}}{T_{\rm x}}
\label{n}
\end{equation}
where the number density of OH, $n_{\rm OH}$, is in m$^{-3}$ and the
excitation temperature $T_{\rm x}$ is --10 K assuming 1 per cent
inversion.  This gives an average $n_{\rm OH}\approx500$ m$^{-3}$ for
$\tau_{\rm u}=-0.4$, and an OH column density of $1.5\times10^{19}$
m$^{-2}$ in 1-pc masing clouds.  CO observations \cite{Bryant96} show
an average molecular column density of $10^{27}$ m$^{-2}$ within 500
mas of the nucleus.  If the OH and the CO are cospatial the fractional
OH number density is $1.5\times10^{-8}$ and the clouds have a
molecular number density of $n\approx3\times10^{10}$ m$^{-3}$. The
fractional abundance of OH could be even higher if the CO extends to a
greater depth than the OH which we observe.

The 85 K dust temperature obtained by \scite{Soifer00} and
\scite{Roche93} produces too high a value of $R_{\rm H}$ for
$n\approx3\times10^{10}$ m$^{-3}$, in the standard model of
\scite{Randell95}.  The models do, however, show a trend towards lower
$R_{\rm H}$ for smaller clouds and/or a larger velocity gradient so
our results might be compatible with our preferred $\sim1$-pc
clouds. The biggest discrepancy arises from our estimated $n_{\rm
OH}\approx500$ m$^{-3}$, which is twice the highest value considered
by \scite{Randell95}. If the OH maser
pump in Mrk 231 is extremely effective with respect to the rate of
stimulated emission (consistent with the
unsaturated maser behaviour), a lower  $n_{\rm OH}$ is possible;
 for example an inversion as high as 10 per
cent would reduce $n_{\rm OH}$ to $\approx50$ m$^{-3}$.  


\subsection{Maser efficiency}
\label{efficiency}

The efficiency $\eta$ of a radiative maser pump is the product of the
ratio of the OH maser to IR pump photon rates and the the line profile
ratio.  IR and optical data including polarimetry on various scales
(\pcite{Soifer00}; \pcite{Siebenmorgen01}; \pcite{Smith04}; see
Section~\ref{polarization}) suggest that the IR emission within the
central arcsec has two distinct origins; a polarized MIR source close
to the core and a cooler 60-$\mu$m source at up to a few hundred pc,
of mainly starburst origin, providing the maser pump.

The main IR lines exciting the maser population inversion are
at 35 and 53 $\mu$m.  The contribution of the 35-$\mu$m line was
demonstrated by absorption measurements of the megamaser galaxy Arp
220 \cite{Skinner97} in relation to the 1667-MHz maser luminosity. The
photon rate ratio between frequencies $ \nu_{\rm OH}$, $\nu_{\rm IR}$,
is given by $\dot{\Phi}_{\rm OH}/\dot{\Phi}_{\rm IR} = \Omega S_{\rm
OH} \nu_{\rm IR}/4 \pi S_{\rm IR} \nu_{\rm OH}$, where $S_{\rm OH}$
and $S_{\rm IR}$ are the flux densities and $\Omega$ is the maser
beaming angle.  We assume that the IR emission covers 4$\pi$ Sr as
\scite{Soifer00} have shown that emission in this range comes from a
region comparable in size to the maser region of Mrk 231
(Section~\ref{polarization}).  The line profile ratio is equivalent to
$\Delta\nu_{\rm OH}/\Delta\nu_{\rm IR} = [\nu_{\rm
OH}(\Delta{V}/c)]/[\nu_{\rm IR}(\Delta{V}/c)]$ where $\Delta{V}$ is
the OH thermal line width.  Combining these expressions and cancelling
common terms gives 

\begin{equation}
\eta \approx \frac{\dot{\Phi}_{\rm OH}}{\dot{\Phi}_{\rm IR} }\frac{\Delta\nu_{\rm OH}}{\Delta\nu_{\rm IR}}
 \approx
\frac{\Omega S_{\rm OH}}{4 \pi S_{\rm IR}}
\label{pump}
\end{equation}

 We
extrapolate between the IRAS measurements of Mrk 231 at 25 and 60
$\mu$m given by \scite{Soifer00} to estimate that, at $\nu_{\rm IR} =
8.6$ THz (corresponding to 35 $\mu$m), $S_{\rm IR}=15.9$ Jy with
$\approx10$ per cent accuracy.  The OH maser peak $S_{\rm OH}=0.038$
Jy at $\nu_{\rm OH}=1667$ MHz. The maser emission from Mrk 231 almost
certainly covers most of $4\pi$ Sr and might even appear brighter to
an observer looking at the torus edge-on (see
Section~\ref{comparisons}).  This leads to $\eta\approx2\times10^{-4}$
which is indeed low for maser models such as those of
\scite{Randell95} who suggest that $\eta$ lies between 0.001 -- 0.01.
The low efficiency and the maximum gain constrained by $R_{\rm
M/C}\le2.2$ (Section~\ref{gain}) are consistent with unsaturated maser
amplification due to a population inversion provided by IR radiation
from dust \cite{Baan85}.  In other
words, the maser population inversion is probably high but only
absorbs a tiny fraction of the IR radiation produced by this ULIRG.

\section{Mrk 231 among other megamaser galaxies}
\label{comparisons}

Mrk 231 is the only known Seyfert 1 which supports OH megamasers.
 They are distributed in a torus which is more nearly face-on (see
 Section~\ref{dynamic}) than the distributions in other megamaser
 galaxies with Seyfert 2 characteristics. The OH masers in Mrk 273,
 III Zw 35 and Arp 220, have been modelled as discs or tori rotating
 about axes at inclinations of $>45$\deg to the line of sight
 (\pcite{Yates00}; \pcite{Pihlstroem01}; \pcite{Rovilos03},
 respectively).

 Mrk 273 is classed as a Seyfert 2 and has an HI absorption column
density of $(17-18)\times10^{23}$ m$^{-2}$ $\times$ (spin
temperature/filling factor), almost three times greater than estimates
for Mrk 231, assuming the same conditions (\pcite{Carilli98s};
\pcite{Cole99}).
MERLIN observations \cite{Yates00} show that the brightest maser peak
in Mrk 273 is 67 mJy~beam$^{-1}$, slightly offset from the 12
mJy~beam$^{-1}$, 1.6-GHz continuum peak.  The maser gain is much
greater than in Mrk 231, suggesting amplification along a longer path
length in Mrk 273, through an edge-on torus.  The 1667:1665-MHz peak
ratio in Mrk 273 is $\ge5$. \scite{Klockner04} measure a lower ratio
on larger scales (e.g. WSRT) and a higher ratio on smaller EVN scales,
but even the most compact masers in Mrk 273 are still unsaturated.
Using the method outlined in Section~\ref{efficiency} gives
$\eta\approx8\times10^{-4}\times\Omega/4\pi$ for the pumping
efficiency in Mrk 273, higher than in Mrk 231 by a factor of  4$\Omega$,
which may be due to a combination of a smaller beaming angle $\Omega$
and better maser amplification (which can be achieved without
saturation if clouds overlap along the line of sight).
\scite{Klockner04} estimate an enclosed compact mass of
$(1.4\pm0.2)\times10^9$ M$_{\odot}$, larger than the value of
$(6\pm1)\times10^8$ M$_{\odot}$ measured on MERLIN scales
(\pcite{Yates00}; \pcite{Richards01})\footnote{Note that the method is
fully explained in \scite{Yates00} but an arithmetic slip leads to the
wrong result; the correct value is given in \scite{Richards01}.} which
suggests that Mrk 273 also contains a massive compact core determining
the maser kinematics on the smallest scales.
 Moreover, Arp 220
\cite{Lonsdale98}, III Zw 35 and IRAS 17208-0014
\cite{Diamond99} all have indications of tori closer to edge-on
and higher maser gains and/or hyperfine ratios than Mrk 231.
 
  The classical relation between OH megamaser, radio continuum and IR
luminosities \cite{Baan89} is given by 
\begin{equation}
L_{\rm M}\propto L_{\rm C}\times L_{\rm
IR} \propto L_{\rm IR}^{\gamma} \propto L_{\rm C}^{\gamma}
\label{quad}
\end{equation}
where $L_{\rm M}$, $L_{\rm C}$ and $L_{\rm IR}$ are the maser, radio
and continuum luminosities respectively and $\gamma=2$.  This rests on
the assumptions that the masers are in a position to amplify the radio
continuum, that they are unsaturated and that the radio continuum and
IR flux densities have a linear relationship.  These are justified if
the dominant sources of all three types of radiation have a common
origin in starburst activity, but not if an AGN or jets are
significant radio sources.  \scite{Kandalian97} obtain $\gamma = 1.38$
(1.66) with (without) taking the Malmquist bias into account, whilst
\scite{Klockner04t} find $0.99< \gamma < 2.29$ for various samples,
with a wide range even for a more complete nearby sample or by using
other completeness qualifiers.  \scite{Darling02}, using high
signal-to-noise OH profiles, find that $L_{\rm M}$, $L_{\rm C}$ and
$L_{\rm IR}$ are increasingly poorly correlated for fainter
sub-samples, arriving at a best value of $\gamma = 1.2\pm0.1$. This
could be due to near-saturation of most OH megamasers (but our results
suggest otherwise), or because the single pan-galactic values of
$L_{\rm C}$ and $L_{\rm IR}$ usually employed mask local variations in
emissivity. Such intrinsic differences are very probable in ULIRGs
viewed at various times since the original mergers apparently
responsible.  There is another possibility, suggested by
\scite{Klockner04t}, which is simply orientation.  This is clearly
important in explaining the maser differences between Mrk 231 and Mrk
273 and other edge-on systems, as it is in explaining their Seyfert
characteristics.

 The status of
an AGN in the maser region of Mrk 273 is contentious but the central
mass condensation ($6-14\times10^8$ M$_{\odot}$, \pcite{Yates00};
\pcite{Klockner04}) appears to be $\sim2$ orders of magnitude larger than that
in Mrk 231.  If galaxy mergers are responsible for funnelling material
into the cores, the initial core mass or the time since merger must be
greater for Mrk 273.
In the latter case, the starburst activity in both galaxies suggests
that it preceeds the emergence of an AGN but also outlives obvious
nuclear activity.

\section{Conclusions}
\label{conclusions}

MERLIN detects and resolves all the OH mainline maser emission and
1.6-GHz radio continuum within several hundred pc of the Seyfert 1
nucleus of Mrk 231. The continuum emission declines fairly smoothly in
all directions from a central peak but is more extended to the S.  It
is 0.4 per cent polarized and the polarization angle implies that the
associated magnetic field runs N-S, similar to the jet direction on
MERLIN scales e.g. \scite{Taylor99}.  Most of the extended radio
emission detected by MERLIN is probably of starburst origin but we see
a region of apparently low maser gain running S which might be due
higher foreground radio continuum from the approaching jet .

 The maser distribution shows a $V_{\rm LSR}$ gradient of
($1.7\pm0.2$) km~s$^{-1}$ pc$^{-1}$ along the 420-pc major axis from
SE to NW. The NE and SW contain extended regions at intermediate
velocities.  This is compelling evidence for a molecular region
rotating about an axis at $i\approx45$\deg at a position angle of
$\approx230$\deg (Section~\ref{sub-kpc}).  The maser kinematics
suggest that the torus contains a mass density of $320\pm90$
M$_{\odot}$ pc$^{-3}$.  Comparison with the EVN results for the
innermost 60 pc \cite{Klockners03} shows that the most likely
configuration at $50-200$ pc is solid body rotation of a flattened
mass distribution.  VLBI observations (\pcite{Carilli98s};
\pcite{Klockners03}) suggest that the inner rim of the torus is at
$4<r_{\rm i}<16$ pc.  The MERLIN data alone reveal anomalously high
velocity emission within 50 pc of the core. If this is at $r_{\rm i}$
and arises from material in Keplerian orbit around the central black
hole, a mass of $\la8\times10^6$ M$_{\odot}$ is implied, making it one
of the lightest active black holes yet weighed.

The orientation of the axis of the maser torus is consistent with EVN
data but misaligned with the radio jets on comparable scales and with
the axes of HI and larger-scale CO rotation, all of which are
projected in roughly N--S directions (Section~\ref{sb_jet}). It is
likely that the molecular torus is warped and it is also possible for
different species to follow different trajectories if neutrals, ions
and stars are influenced differently in a barred potential.

  The maser intensity as a function of position is not closely
correlated with the 1.6-GHz continuum. The apparent optical depth is
consistently close to the unsaturated maser optical depth.  The maser
gain with respect to the radio continuum is $\le2.2$.  The average
1667:1665-MHz hyperfine emission ratio is the thermal equillibrium
value of 1.8.  The fraction of maser emission detected from both
mainline transitions remains  proportional to the
resolution over 4 orders of magnitude, from single-dish (9 arcmin,
\pcite{Stavely-Smith87}) through WSRT, MERLIN and EVN resolutions (14
arcsec to 40 mas, this paper and \pcite{Klockners03}) to global VLBI
(5 mas, \pcite{Lonsdale03}).  These properties imply that the maser
emission is unsaturated with a negative optical depth less than one
\cite{Henkel90}, from clouds with a covering factor close to unity.

The kinematics of the torus suggest that the maser emission can be
amplified over a velocity coherent path length of $\ge0.25$ pc, see
Section~\ref{clouds}. The peak maser brightness temperature appears to
be $(2.2-2.5)\times10^6$ K for the 1667-MHz line for all spatially
resolved observations.  This suggests that the maser clouds have a
consistent distribution and filling factor on scales from the global
VLBI resolution (4 pc) up to the MERLIN resolution of 100 pc, and that
they are smaller than the finest resolution. These results imply cloud
sizes within a factor of four of 1 pc. The models of
\scite{Randell95} give a number density of $\approx3\times10^{10}$
m$^{-3}$ with $n_{\rm OH}\approx50$ m$^{-3}$ for a 10 per cent
population inversion in the masing transitions. The efficiency with
which MIR radiation \cite{Soifer00} is converted to maser emission is
low, $\approx2\times10^{-4}$ (Section~\ref{efficiency}), so only
radiative pumping \cite{Baan85} is required.

These results show conclusively that there are no very high surface
brightness 1667-MHz masers in Mrk 231, which can be explained if they
are located in a torus which is thinner in the tangential direction
than in the radial direction (with respect to the galactic dynamic
centre).  From our viewpoint, the more face-on orientation of a
Seyfert 1 like Mrk 231 produces a shorter maser amplification depth
and lower gain than occurs in megamaser galaxies such as Mrk 273 which
have Seyfert 2 characteristics, as discussed in
Section~\ref{comparisons}. The unified scheme model \cite{Antonucci85}
explains the optical characteristics of Seyfert galaxies in terms of
the orientation of an inner pc-scale torus.  We find that, where an
associated larger-scale torus containing starburst activity is traced
by megamasers, it appears that orientation can also explain some of the
variation in apparent maser intensity relative to the radio continuum.

\section{Acknowledgements}

MERLIN is the multi-element radio linked interferometer network,
operated by the University of Manchester on behalf of PPARC. We warmly
thank Drs. Alan Pedlar and Rob Beswick and Professors Phil Diamond and
Robert Laing for useful discussions.  We are very grateful to
Dr. Pedlar for encouraging our use of his MERLIN archive continuum
data and to Drs. Peter Thomasson, Tom Muxlow and Simon Garrington and
the other MERLIN staff for assistance with the observations and data
reduction. We have made use of observations made with the NASA/ESA
{\em Hubble Space Telescope}, obtained from the data archive at the Space
Telescope Science Institute.  STScI is operated by the Association of
Universities for Research in Astronomy, Inc. under NASA contract NAS
5-26555.

We thank the anonymous referee for very perceptive and helpful
comments which have added depth to this paper.
We acknowledge the use of data reduction software and
hardware provided by the PPARC STARLINK project and of the services
provided by NASA ADS and by CDS.  AMSR thanks the AVO and AstroGrid
for funding during this project and has made use of their Virtual
Observatory tools.

\bibliographystyle{mnras}


\vspace*{2cm}
\appendix
\section{}
\begin{table}
\caption{The properties of OH mainline maser components in Mrk
  231. ($x$, $y$) positions are relative to the reference position
  12\h~56\m14\fs~2383 +56\degr~52\arcmin~25\farcs210 (J2000).}
\label{M231NEWSPOTS.TAB}
\begin{tabular}{lcccrrrr}
\hline
Reg.&Freq.&$V_{\rm hel}$ & $V_{\rm LSR}$&$x$&y&$\sigma_{xy}$&$P$\\
    &(MHz)&\multicolumn{2}{c}{(km s$^{-1}$)} &\multicolumn{3}{c}{(mas)}&(mJy \\
&&&&&&& b$^{-1}$)\\
(1)&(2)&(3)&(4)&(5)&(6)&(7)&(8)\\
\hline
\multicolumn{8}{l}{Identified with 1665 MHz line}\\
 SE & 1596.367 & 12954& 12427 & 236 & --142 &  96 &   3\\
 SE & 1596.492 & 12930& 12405 &  88 & --153 &  77 &   3\\
 SE & 1596.617 & 12905& 12382 & 154 & --104 &  91 &   2\\
 SE & 1596.742 & 12881& 12360 &  97 & --105 & 112 &   3\\
 SE & 1596.867 & 12856& 12337 & 147 & --154 &  79 &   3\\
    &	       &      &									&     &	  &	&\\
  C & 1596.367 & 12954& 12427 &   0 &   53 & 120 &   2\\
  C & 1596.492 & 12930& 12405 &   8 &   23 &  87 &   2\\
  C & 1596.617 & 12905& 12382 &  19 &   16 &  63 &   4\\
  C & 1596.742 & 12881& 12360 &  34 &  --11 &  44 &   5\\
  C & 1596.867 & 12856& 12337 &   9 &   23 &  39 &   6\\
  C & 1596.992 & 12832& 12315 &  --5 &   34 &  31 &   7\\
  C & 1597.117 & 12807& 12292 &  --7 &   35 &  29 &   8\\
  C & 1597.242 & 12783& 12270 &  --6 &    6 &  24 &   9\\
  C & 1597.367 & 12758& 12247 &  --6 &   29 &  19 &  11\\
  C & 1597.492 & 12734& 12225 &  --1 &   28 &  17 &  12\\
  C & 1597.617 & 12709& 12202 & --18 &   26 &  18 &  11\\
  C & 1597.742 & 12685& 12180 & --24 &   37 &  12 &  17\\
  C & 1597.867 & 12660& 12157 & --28 &   50 &  12 &  16\\
  C & 1597.992 & 12636& 12135 & --14 &   56 &  11 &  17\\
  C & 1598.117 & 12611& 12112 & --29 &   64 &  13 &  14\\
  C & 1598.242 & 12588& 12090 & --16 &   64 &  14 &  14\\
  C & 1598.367 & 12563& 12067 & --30 &   68 &  20 &  10\\
    &	       &      &									&     &	  &	&\\
 SW & 1597.492 & 12734& 12225 &--195 &  --31 &  59 &   3\\
 SW & 1597.617 & 12709& 12202 &--113 &   30 & 105 &   3\\
 SW & 1597.742 & 12685& 12180 &--122 &  --43 &  55 &   4\\
 SW & 1597.867 & 12660& 12157 &--145 &   --7 &  41 &   5\\
 SW & 1597.992 & 12636& 12135 &--131 &  --12 &  67 &   3\\
 SW & 1598.117 & 12611& 12112 &--138 &    9 &  54 &   4\\
 SW & 1598.242 & 12588& 12090 &--133 &   77 &  83 &   2\\
 SW & 1598.992 & 12441& 11955 & --96 &  --89 &  55 &   3\\
 SW & 1599.117 & 12416& 11932 & --92 &  --48 &  63 &   3\\
 SW & 1599.242 & 12392& 11910 & --28 &  --91 &  55 &   3\\
 SW & 1599.367 & 12367& 11887 & --57 & --106 &  59 &   3\\
    &	       &      &									&     &	  &	&\\
 NE & 1598.742 & 12490& 12000 &  73 &   66 &  68 &   3\\
 NE & 1598.867 & 12465& 11977 &  58 &    4 &  71 &   3\\
 NE & 1598.992 & 12441& 11955 &  48 &   42 &  88 &   3\\
 NE & 1599.117 & 12416& 11932 &  86 &   53 &  65 &   3\\
 NE & 1599.242 & 12392& 11910 & 106 &   18 &  61 &   3\\
 NE & 1599.367 & 12367& 11887 &  92 &   21 &  49 &   4\\
 NE & 1599.492 & 12343& 11865 & 103 &   --1 &  46 &   4\\
 NE & 1599.617 & 12318& 11842 &  92 &  --30 &  34 &   5\\
 NE & 1599.742 & 12295& 11820 &  75 &  --11 &  82 &   3\\
    &	       &      &									&     &	  &	&\\
 NW & 1598.742 & 12490& 12000 & --33 &  158 &  75 &   3\\
 NW & 1598.867 & 12465& 11977 & --80 &  136 &  71 &   3\\
 NW & 1598.992 & 12441& 11955 &--135 &  117 &  53 &   3\\
 NW & 1599.117 & 12416& 11932 &--105 &  145 &  59 &   3\\
 NW & 1599.242 & 12392& 11910 & --19 &  169 &  43 &   3\\
 NW & 1599.492 & 12343& 11865 &  41 &  151 &  70 &   3\\
 NW & 1599.617 & 12318& 11842 & --24 &  169 &  56 &   3\\
 NW & 1599.867 & 12270& 11797 & --73 &  181 &  70 &   3\\
 NW & 1599.992 & 12246& 11775 & --89 &  187 &  76 &   3\\
 NW & 1600.117 & 12221& 11752 & --87 &  172 &  85 &   2\\
 NW & 1600.242 & 12197& 11730 &--127 &  160 &  65 &   3\\
 NW & 1600.367 & 12172& 11707 &--152 &  209 &  49 &   3\\
\hline
\end{tabular}
\end{table}
\begin{table}
\contcaption{Maser components in Mrk 231.}
\begin{tabular}{lcccrrrr}
\hline
Reg.&Freq.&$V_{\rm hel}$ & $V_{\rm LSR}$&$x$&y&$\sigma_{xy}$&$P$\\
    &(MHz)&\multicolumn{2}{c}{(km s$^{-1}$)} &\multicolumn{3}{c}{(mas)}&(mJy \\
&&&&&&& b$^{-1}$)\\
(1)&(2)&(3)&(4)&(5)&(6)&(7)&(8)\\
\hline
\multicolumn{8}{l}{Identified with 1667 MHz line}\\
 SE & 1598.367 &12930  & 12405 &  35 &--154 &  64 &   3\\
 SE & 1598.492 &12906  & 12383 &  21 & --72 &  49 &   4\\
 SE & 1598.617 &12881  & 12360 &  56 & --64 &  45 &   4\\
 SE & 1598.742 &12857  & 12338 &  24 & --94 &  47 &   4\\
 SE & 1598.867 &12832  & 12315 &   8 & --80 &  71 &   3\\
    &	       &       &	&     &	  &	&\\
  C & 1598.492 &12906  & 12383 & --29 &  51 &  15 &  12\\
  C & 1598.617 &12881  & 12360 & --15 &  46 &  13 &  14\\
  C & 1598.742 &12857  & 12338 & --21 &  30 &  11 &  17\\
  C & 1598.867 &12832  & 12315 & --21 &  24 &  11 &  18\\
  C & 1598.992 &12808  & 12293 & --23 &  20 &  10 &  18\\
  C & 1599.117 &12783  & 12270 & --19 &  21 &   9 &  21\\
  C & 1599.242 &12759  & 12248 & --23 &  31 &   8 &  23\\
  C & 1599.367 &12734  & 12225 & --23 &  34 &   7 &  27\\
  C & 1599.492 &12710  & 12203 & --24 &  39 &   7 &  27\\
  C & 1599.617 &12685  & 12180 & --24 &  47 &   6 &  31\\
  C & 1599.742 &12661  & 12158 & --25 &  55 &   5 &  35\\
  C & 1599.867 &12636  & 12135 & --28 &  62 &   5 &  37\\
  C & 1599.992 &12613  & 12113 & --29 &  66 &   5 &  38\\
  C & 1600.117 &12588  & 12090 & --36 &  69 &   6 &  32\\
  C & 1600.242 &12564  & 12068 & --34 &  69 &   7 &  30\\
  C & 1600.367 &12540  & 12046 & --42 &  71 &   9 &  23\\
  C & 1600.492 &12515  & 12023 & --44 &  65 &  12 &  17\\
  C & 1600.617 &12491  & 12001 & --44 &  55 &  16 &  13\\
  C & 1600.742 &12466  & 11978 & --46 &  63 &  19 &  11\\
  C & 1600.867 &12442  & 11956 & --67 &  54 &  18 &  12\\
  C & 1600.992 &12417  & 11933 & --47 &  48 &  24 &   9\\
  C & 1601.117 &12393  & 11911 & --50 &  60 &  40 &   6\\
  C & 1601.242 &12368  & 11888 & --78 &  50 &  48 &   5\\
  C & 1601.367 &12345  & 11866 & --87 &  58 &  64 &   4\\
  C & 1601.492 &12320  & 11843 & --87 &  51 &  38 &   7\\
  C & 1601.617 &12296  & 11821 &  --1 &  29 &  78 &   3\\
  C & 1601.742 &12271  & 11798 &  10 &  61 &  65 &   4\\
  C & 1601.867 &12247  & 11776 & --21 &  78 &  66 &   4\\
  C & 1601.992 &12222  & 11753 & --10 &  45 &  96 &   3\\
  C & 1602.117 &12198  & 11731 & --29 &  58 &  51 &   5\\
  C & 1602.242 &12173  & 11708 & --13 &  62 &  64 &   4\\
 \hline
\end{tabular}
\end{table}
\begin{table}
\contcaption{Maser components in Mrk 231.}

\begin{tabular}{lcccrrrr}
\hline
Reg.&Freq.&$V_{\rm hel}$ & $V_{\rm LSR}$&$x$&y&$\sigma_{xy}$&$P$\\
    &(MHz)&\multicolumn{2}{c}{(km s$^{-1}$)} &\multicolumn{3}{c}{(mas)}&(mJy \\
&&&&&&& b$^{-1}$)\\
(1)&(2)&(3)&(4)&(5)&(6)&(7)&(8)\\
\hline
\multicolumn{8}{l}{Identified with 1667 MHz line}\\
 SW & 1599.367 & 12734& 12225 &--146 &  21 &  47 &   4\\
 SW & 1599.492 & 12710& 12203 &--116 & --35 &  51 &   4\\
 SW & 1599.617 & 12685& 12180 &--111 & --21 &  39 &   5\\
 SW & 1599.742 & 12661& 12158 &--118 & --27 &  23 &   8\\
 SW & 1599.867 & 12636& 12135 &--123 & --19 &  39 &   6\\
 SW & 1599.992 & 12613& 12113 &--120 & --50 &  27 &   7\\
 SW & 1600.117 & 12588& 12090 &--138 & --31 &  56 &   4\\
 SW & 1600.242 & 12564& 12068 &--123 & --12 &  44 &   5\\
 SW & 1600.367 & 12540& 12046 &--127 & --21 &  39 &   5\\
 SW & 1600.492 & 12515& 12023 &--141 &  --9 &  44 &   5\\
 SW & 1600.617 & 12491& 12001 & --98 & --66 &  46 &   4\\
 SW & 1600.742 & 12466& 11978 &--117 & --51 &  45 &   5\\
 SW & 1600.867 & 12442& 11956 &--105 & --67 & 101 &   2\\
 SW & 1600.992 & 12417& 11933 &--101 & --59 &  75 &   3\\
    &	       &      &	&     &	  &	&\\
 NE & 1599.742 & 12661& 12158 &  75 &  98 &  65 &   4\\
 NE & 1599.867 & 12636& 12135 &  74 & 106 &  35 &   5\\
 NE & 1599.992 & 12613& 12113 &  95 &  71 &  39 &   5\\
 NE & 1600.117 & 12588& 12090 &  64 &  33 &  65 &   3\\
 NE & 1600.242 & 12564& 12068 &  65 & 122 &  40 &   5\\
 NE & 1600.367 & 12540& 12046 &  70 &  87 &  45 &   4\\
 NE & 1600.492 & 12515& 12023 &  62 &  91 &  63 &   3\\
 NE & 1600.617 & 12491& 12001 &  38 & 128 &  67 &   3\\
 NE & 1600.742 & 12466& 11978 & 117 & --15 &  62 &   3\\
 NE & 1600.867 & 12442& 11956 &  67 &  55 &  73 &   3\\
 NE & 1600.992 & 12417& 11933 &  38 & 126 &  47 &   5\\
 NE & 1601.117 & 12393& 11911 &  65 & 141 &  49 &   5\\
 NE & 1601.242 & 12368& 11888 &  40 & 106 &  63 &   4\\
    &	       &      &	&     &	  &	&\\
 NW & 1600.617 & 12491& 12001 & --65 & 268 &  67 &   3\\
 NW & 1600.742 & 12466& 11978 &--137 & 245 &  62 &   3\\
 NW & 1600.867 & 12442& 11956 &--107 & 235 &  73 &   3\\
 NW & 1601.117 & 12393& 11911 &--164 & 201 &  77 &   3\\
 NW & 1601.367 & 12345& 11866 &--145 & 188 & 114 &   2\\
 NW & 1601.492 & 12320& 11843 &   7 & 151 &  98 &   3\\
 NW & 1601.617 & 12296& 11821 & --51 & 289 &  58 &   5\\
 NW & 1601.742 & 12271& 11798 &--108 & 262 &  96 &   3\\
\hline
\end{tabular}
\end{table}

\end{document}